\documentclass[aps,prl,reprint,amsmath,amssymb,superscriptaddress,showpacs]{revtex4-2}
\usepackage{graphicx}
\usepackage{dcolumn}
\usepackage{bm}
\usepackage[colorlinks,linkcolor=blue,anchorcolor=blue,citecolor=blue,urlcolor=blue,filecolor=blue,menucolor=blue,runcolor=blue]{hyperref}
\par
\begin{document}
\title{Revealing the Heavy Quasiparticles in the Heavy-Fermion Superconductor CeCu$_2$Si$_2$}
\author{Zhongzheng Wu}
\affiliation{Center for Correlated Matter and Department of Physics, Zhejiang University, Hangzhou, China}
\author{Yuan Fang}
\affiliation{Center for Correlated Matter and Department of Physics, Zhejiang University, Hangzhou, China}
\author{Hang Su}
\affiliation{Center for Correlated Matter and Department of Physics, Zhejiang University, Hangzhou, China}
\author{Wu Xie}
\affiliation{Center for Correlated Matter and Department of Physics, Zhejiang University, Hangzhou, China}
\author{Peng Li}
\affiliation{Center for Correlated Matter and Department of Physics, Zhejiang University, Hangzhou, China}
\author{Yi Wu}
\affiliation{Center for Correlated Matter and Department of Physics, Zhejiang University, Hangzhou, China}
\author{Yaobo Huang}
\affiliation{Shanghai Institute of Applied Physics, Chinese Academy of Science, Shanghai, China}
\author{Dawei Shen}
\affiliation{State Key Laboratory of Functional Materials for Informatics and Center for Excellence in Superconducting Electronics, SIMIT, Chinese Academy of Science, Shanghai, China}
\author{Balasubramanian Thiagarajan}
\affiliation{MAX IV Laboratory, Lund University, P.O. Box 118, 221 00 Lund, Sweden}
\author{Johan Adell}
\affiliation{MAX IV Laboratory, Lund University, P.O. Box 118, 221 00 Lund, Sweden}
\author{Chao Cao}
\affiliation{Department of Physics, Hangzhou Normal University, Hangzhou, China}
\author{Huiqiu Yuan}
\affiliation{Center for Correlated Matter and Department of Physics, Zhejiang University, Hangzhou, China}
\affiliation{Zhejiang Province Key Laboratory of Quantum Technology and Device, Zhejiang University, Hangzhou, China}
\affiliation{Collaborative Innovation Center of Advanced Microstructures, Nanjing University, Nanjing, China}
\author{Frank Steglich}
\affiliation{Center for Correlated Matter and Department of Physics, Zhejiang University, Hangzhou, China}
\affiliation{Max Planck Institute for Chemical Physics of Solids, 01187 Dresden, Germany}
\author{Yang Liu}%
\email {yangliuphys@zju.edu.cn}
\affiliation{Center for Correlated Matter and Department of Physics, Zhejiang University, Hangzhou, China}
\affiliation{Zhejiang Province Key Laboratory of Quantum Technology and Device, Zhejiang University, Hangzhou, China}
\affiliation{Collaborative Innovation Center of Advanced Microstructures, Nanjing University, Nanjing, China}
\date{\today}%
\addcontentsline{toc}{chapter}{Abstract}

\begin{abstract}
The superconducting order parameter of the first heavy-fermion superconductor CeCu$_2$Si$_2$ is currently under debate. A key ingredient to understand its superconductivity and physical properties is the quasiparticle dispersion and Fermi surface, which remains elusive experimentally. Here we present measurements from angle-resolved photoemission spectroscopy. Our results emphasize the key role played by the Ce $4f$ electrons for the low-temperature Fermi surface, highlighting a band-dependent conduction-$f$ electron hybridization. In particular, we find a very heavy quasi-two-dimensional electron band near the bulk \textit{X} point and moderately heavy three-dimensional hole pockets near the \textit{Z} point. Comparison with theoretical calculations reveals the strong local correlation in this compound, calling for further theoretical studies. Our results provide the electronic basis to understand the heavy fermion behavior and superconductivity; implications for the enigmatic superconductivity of this compound are also discussed.
\end{abstract}

\maketitle

Heavy-fermion (HF) materials are prototypical strongly correlated systems and serve as prime examples for studying quantum criticality and unconventional superconductivity \cite{ref1,ref2,ref3,ref4,ref5,ref6}. CeCu$_2$Si$_2$ is the first discovered HF superconductor \cite{ref7}, and the large size of both the jump in the electronic specific heat \cite{ref7} and the initial slope of the upper critical field curve \cite{ref7a,ref7b} at the superconducting transition temperature $T_C$ $\sim$ 0.6 K indicates that the Cooper pairs are formed by heavy-mass quasiparticles due to the many-body Kondo effect. The unconventional superconductivity in CeCu$_2$Si$_2$ is closely linked to an anti-ferromagnetic quantum critical point, hence very likely driven by the associated quantum fluctuations \cite{ref8,ref10}. Indeed, measurements of inelastic neutron scattering (INS), nuclear quadrupole resonance, and angle-resolved upper critical field indicate that at temperatures above 100 mK, CeCu$_2$Si$_2$ exhibits a nodal $d$-wave superconducting order parameter \cite{ref10,ref12,ref13,ref14,ref15}. However, recent measurements of both the specific heat and London penetration depth hint at a finite gap at very low temperatures \cite{ref16,ref17,ref18}, which stimulates debates on its superconducting order parameter \cite{ref16,ref17,ref18,ref23,ref19,ref21,ref22}. A proposed \textit{d}+\textit{d} band-mixing pairing can well explain the existing experimental results \cite{ref17,ref23,Nica2021npjQM}, although other superconducting pairing symmetries are also proposed, including the $s_\pm$ pairing and $s_{++}$ pairing \cite{ref18,ref19,ref21,ref22}.

The Fermi surface (FS) is vital for understanding the superconductivity and physical properties. However, the FS knowledge that we have so far for CeCu$_2$Si$_2$ is mostly based on theoretical calculations. In particular, the renormalized band calculation (RBC) has predicted the heavy cylindrical band near the bulk \textit{X} point \cite{ref24,ref25}, which nicely explains the ordering vector of the spin density wave (SDW) observed in the neutron scattering \cite{ref10,ref10a}. Similar electron band has been obtained by the density-functional theory (DFT) calculation including Hubbard \textit{U} of $4f$ electrons (DFT + \textit{U}) \cite{ref16,ref18,ref21,ref22}, although the calculated mass is much smaller. Another recent study using the DFT + Gutzwiller method can reproduce this heavy electron band with larger effective mass \cite{ref26}. Dynamic mean-field theory (DMFT) calculations were also employed \cite{ref27,LuoSCPMA2020}, but the resulting FS topology appears to differ from aforementioned methods. Experimentally, quantum oscillations (QOs) had been reported for CeCu$_2$Si$_2$ \cite{HuntJPCM1990,TayamaPRB2003}, which revealed only one orbital with moderately enhanced effective mass ($\sim$5 $m_e$). A previous angle-integrated photoemission study on polycrystalline samples revealed Kondo resonance peaks near the Fermi level ($E_F$) \cite{ref28}, but the momentum-resolved FS was not available. A scanning tunneling spectroscopy (STS) study reported two superconducting gaps \cite{ref29}, which hints at a multigap order parameter. Due to the severe difficulty of sample cleavage for CeCu$_2$Si$_2$, electron spectroscopic measurements that typically require flat cleaved surfaces are very challenging.

Here we report the momentum-resolved measurement of the electronic structure of CeCu$_2$Si$_2$ by angle-resolved photoemission spectroscopy (ARPES), providing direct spectroscopic proof for the multiband FS with strong contributions from Ce 4\textit{f} electrons. The details of single crystal synthesis, characterization and cleavage methods as well as DFT calculations can be found in \cite{ref30,refs41,refs42,refs43,refs44,refs45,refs46,refs47,refs48,ref39,refs49}. Synchrotron ARPES measurements were performed at 10 K or 20 K at beamlines BL03U and BL09U in Shanghai Synchrotron Radiation Facility, and the BLOCH beamline in MAX IV lab, using $p$-polarized light. Our successful ARPES measurement was made possible by improved sample preparation method and recently developed ARPES technique with very small beam spot \cite{ref30}.

\begin{figure}[ht]
\centering
\includegraphics[scale=0.63]{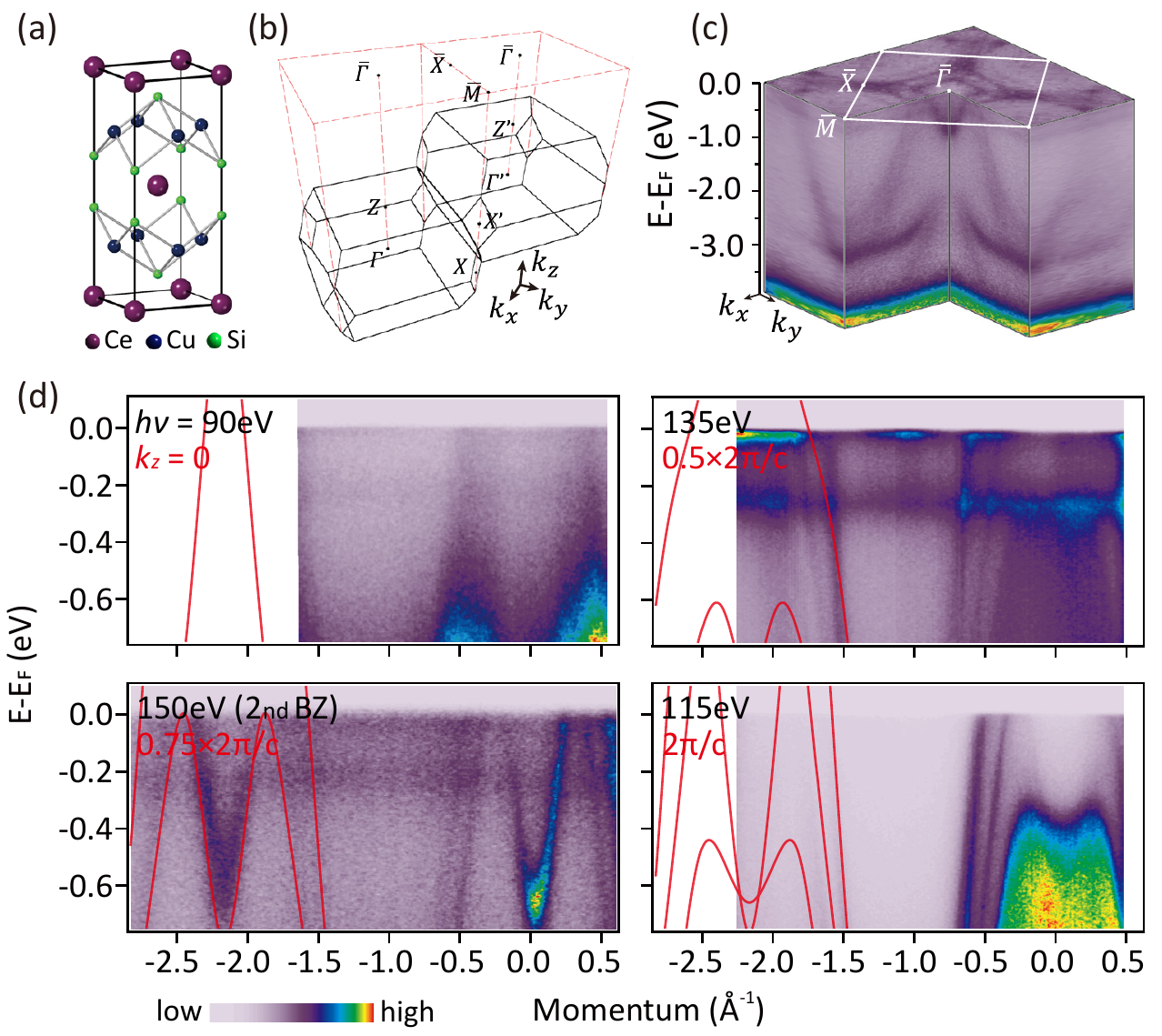}
\caption{(Color Online). Overview of the electronic structure of CeCu$_2$Si$_2$. (a) Crystal structure. (b) 3D bulk BZ (black) and projected surface BZ (red). (c) A 3D view of the electronic structure taken with 150 eV photons from the second surface BZ. (d) ARPES data taken along $\Bar{\varGamma}$-$\bar{M}$ under some representative photon energies, in comparison with results of localized 4\textit{f} calculations. Only the calculation results in the second BZ are shown for clarity. }
\end{figure}

CeCu$_2$Si$_2$ crystallizes in the body-centered tetragonal ThCr$_2$Si$_2$ structure (space group I4/mmm, Fig. 1(a)). Its three-dimensional (3D) bulk Brillouin zone (BZ) and projected surface BZ are shown in Fig. 1(b). The bulk $\varGamma$ ($k_z$=0) and \textit{Z} ($k_z$=2$\pi$/c) points project onto the surface $\Bar{\varGamma}$ point, while the bulk \textit{X} point projects onto the surface $\bar{M}$ point. A 3D ($E$-$k_x$-$k_y$) view of the overall electronic structure is presented in Fig. 1(c): the in-plane $k_x$-$k_y$ map features a few FS pockets centered at $\Bar{\varGamma}$ and $\bar{M}$; the energy-momentum dispersions reveal dispersive conduction bands near $E_F$, the localized 4$f^0$ peak at -2.5 eV, and an intense 3$d$ band at -3.8 eV. A few representative photon energy scans, corresponding to different $k_z$ cuts, are summarized in Fig. 1(d), together with the localized 4\textit{f} calculation (where the 4\textit{f} electrons are treated to be core electrons). Here we used the localized 4\textit{f} calculation for comparison with the overall dispersion of the non-4\textit{f} conduction bands, because the many-body Kondo effect causes distortion of the conduction bands only in the vicinity of $E_F$ via conduction-\textit{f} hybridization (\textit{c}-\textit{f} hybridization) \cite{ref31}. DFT + \textit{U} calculations tend to overestimate the influence of 4\textit{f} electrons for conduction bands, which will be discussed below. A systematic comparison between experiments and calculations yields an estimated inner potential of 11.6 eV \cite{ref30}. The M-shaped conduction bands near $\Bar{\varGamma}$ are quite 3D and exhibit clear $k_z$ dispersion, in good agreement with the localized 4\textit{f} calculation. The spectra taken with \textgreater 122 eV photons (where $4f$-electron cross section becomes appreciable) exhibit flat bands at $\sim$ -0.25 eV and $E_F$, which are the 4$f_{7/2}^1$ and 4$f_{5/2}^1$ Kondo resonance peaks expected for Ce-based Kondo systems \cite{ref28,patil2016arpes,ref33}. The presence of these peaks indicates that Ce 4\textit{f} electrons play an important role in the FS and low-energy excitations.

\begin{figure}[ht]
\centering
\includegraphics[scale=0.56]{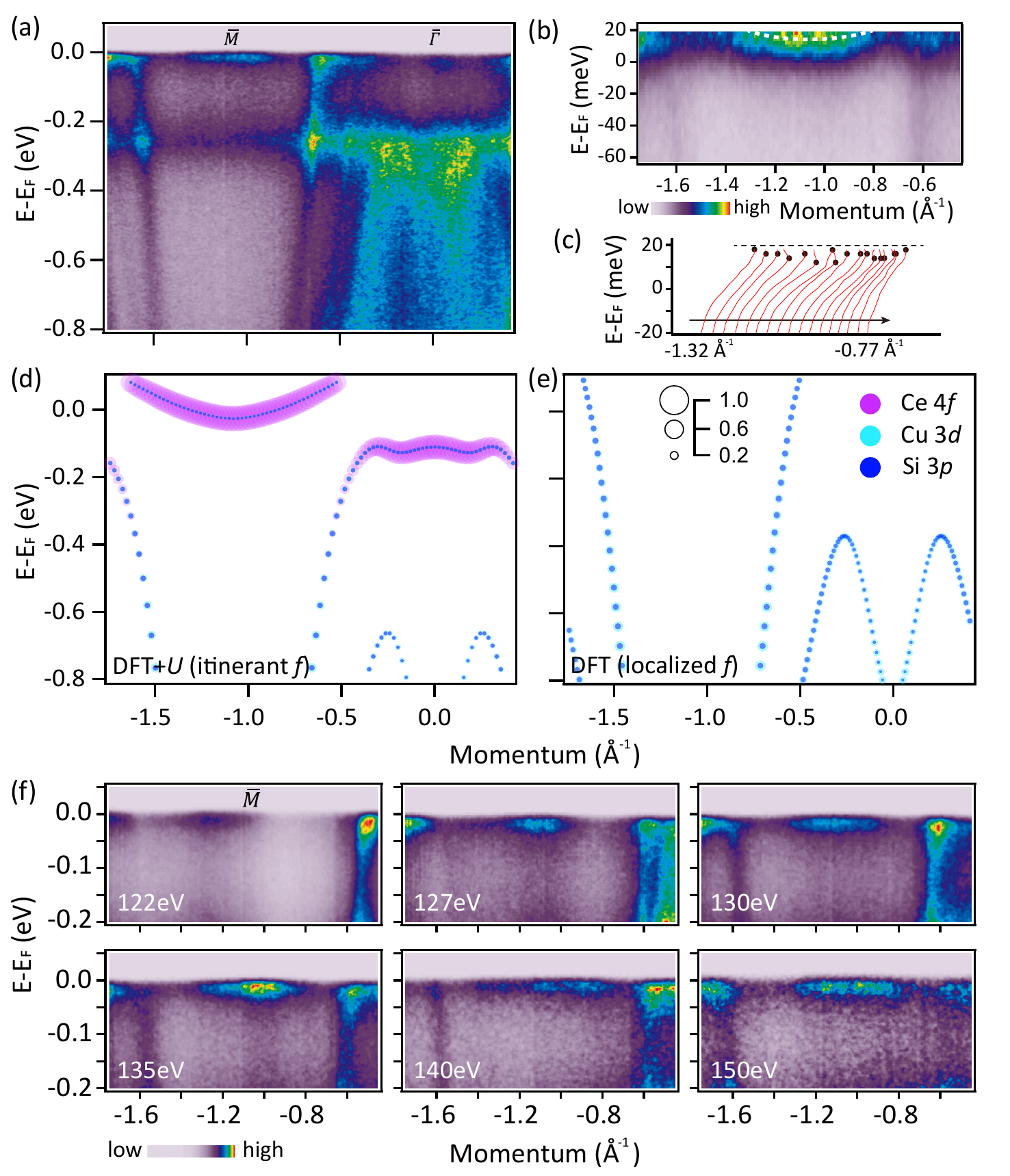}
\caption{(Color Online). The heavy electron band near $\bar{M}$ at 10 K. (a) Band dispersion along $\Bar{\varGamma}$-$\bar{M}$ recorded with 132 eV photons. (b) The spectrum in (a) divided by RC-FDD near $E_F$. (c) Energy distribution curves (EDCs) of (b). The extracted peak positions are highlighted as black dots, which are fitted by a parabola shown as white dashed curve in (b). (d,e) Band dispersions from DFT + \textit{U} (d) and localized 4$f$ calculations (e), for comparison with (a). The orbital contributions are also indicated. (f) Photon-energy dependence of the heavy electron band along $\Bar{\varGamma}$-$\bar{M}$. }
\end{figure}

Previous calculations using RBC and DFT + \textit{U} predict a heavy electron band of mostly $4f$ character at $E_F$ near the bulk \textit{X} point \cite{ref12,ref21,ref22,ref24,ref25}, which is considered of primary importance for the HF superconductivity in CeCu$_2$Si$_2$. Fig. 2(a) shows the quasiparticle dispersion near $E_F$ along $\bar{\varGamma}$-$\bar{M}$, where we can observe a flat electron-type band right at $E_F$ centered at $\bar{M}$ (where the bulk \textit{X} point projects onto). The intensity of this band is maximal near $\bar{M}$ and diminishes away from $\bar{M}$ \cite{ref30}. Such spectral behavior is consistent with a very heavy electron band centered at $\bar{M}$ and sitting very close to $E_F$, whose photoemission intensity is truncated by the Fermi edge. This heavy electron band can be qualitatively explained by the DFT + \textit{U} calculation shown in Fig. 2(d). Note that the DFT + \textit{U} calculation predicts large modification to the conduction band dispersion due to the presence of 4\textit{f} electrons, which is apparently exaggerated when comparing with the experimental data in Fig. 2(a). By contrast, the localized $4f$ calculation gives a reasonable description for the non-$4f$ conduction bands away from $E_F$ (Fig. 2(e)), but it cannot account for the $4f$ bands near $E_F$. To reveal the full 4\textit{f} spectral function near $E_F$, we divide the original ARPES spectra by the resolution-convoluted Fermi-Dirac function (RC-FDD) \cite{ref28,ref34,ref35}, obtained from fitting to an Au reference spectrum taken under identical conditions \cite{ref30}. Such division (Fig. 2(b)) reveals an electron-type band slightly above $E_F$, which is apparently much flatter (hence heavier) than the DFT + \textit{U} calculation in Fig. 2(d). A parabolic fitting of the extracted peak positions (Fig. 2(b,c)) yields an estimated effective mass of $\sim$120 $m_e$. Note that the effective mass estimated from the electronic specific heat ($\sim$0.7 J$\cdot$mol$^{-1}\cdot$K$^{-2}$) and used in RBC is $\sim$500 $m_e$ \cite{ref24,ref25}. Since the effective mass estimation for very heavy bands by ARPES can have relatively large error bar due to limited energy resolution and other complications, the order-of-magnitude agreement between ARPES and the specific heat is fairly reasonable. We mention that Kondo screening involving excited crystal electric field (CEF) states could also give rise to satellite peaks in the $4f$ bands \cite{ref28,ref36,ref37}, although these are not resolved in the current case.

\begin{figure}[ht]
\centering
\includegraphics[scale=0.85]{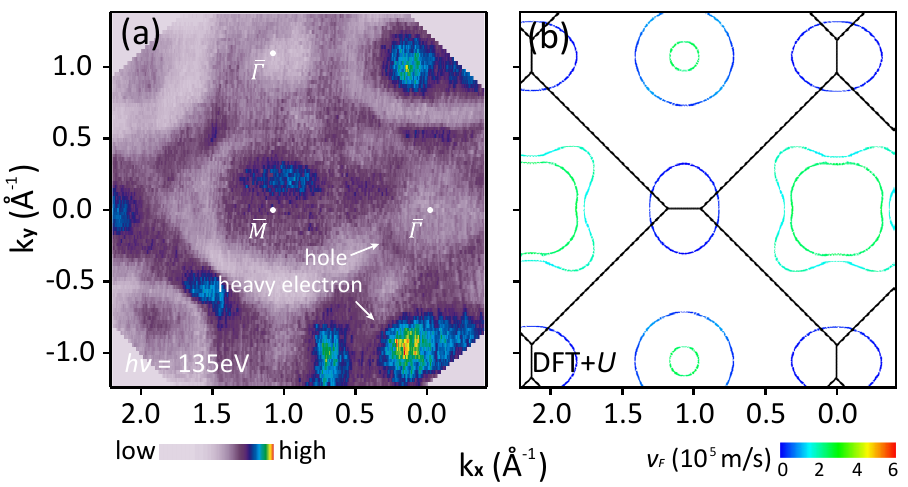}
\caption{(Color Online). (a) The experimental $k_x$-$k_y$ FS at 10 K, taken with 135 eV photons. An energy integration window of 50 meV was used for better statistics. (b) Calculated 2D FS contour from DFT + \textit{U}, for comparison with (a). In (b), the black lines mark the bulk BZ boundaries and the band color indicates the group velocity $v_F$ at $E_F$ (blue indicates heavy band). }
\end{figure}

Fine photon-energy dependent scans, summarized in Fig. 2(f), show that this heavy electron band can be observed in a large $k_z$ range with roughly similar in-plane momentum extent, implying its quasi-two-dimensional (2D) nature. Indeed, both RBC and DFT + \textit{U} calculations predict that this heavy band is quasi-2D with a warped cylindrical shape, exhibiting a nesting vector \textbf{\textit{q}} $\sim$ (0.2, 0.2, 0.5) \cite{ref12,ref21,ref10a}. To further probe its in-plane contour, we measured the $k_x$-$k_y$ map covering a few surface BZ's (Fig. 3(a)). The experimental FS reveals a roughly circular pocket at every $\bar{M}$ point, derived from this heavy electron band, which roughly agrees with the DFT + \textit{U} calculation shown in Fig. 3(b). While both RBC and DFT + \textit{U} calculations predict that this heavy electron band should be slightly elliptical with $k_z$-dependent warping, this is difficult to resolve at the moment due to the broad contour as a result of its large effective mass. In addition to the heavy electron pocket, the experimental FS in Fig. 3(a) also exhibits a small hole pocket centered at $\Bar{\varGamma}$, in reasonable agreement with the calculation, and another large pocket centered at $\bar{M}$, which is not present in the DFT + \textit{U} calculation \cite{ref30}.

\begin{figure}[ht]
\centering
\includegraphics[scale=0.85]{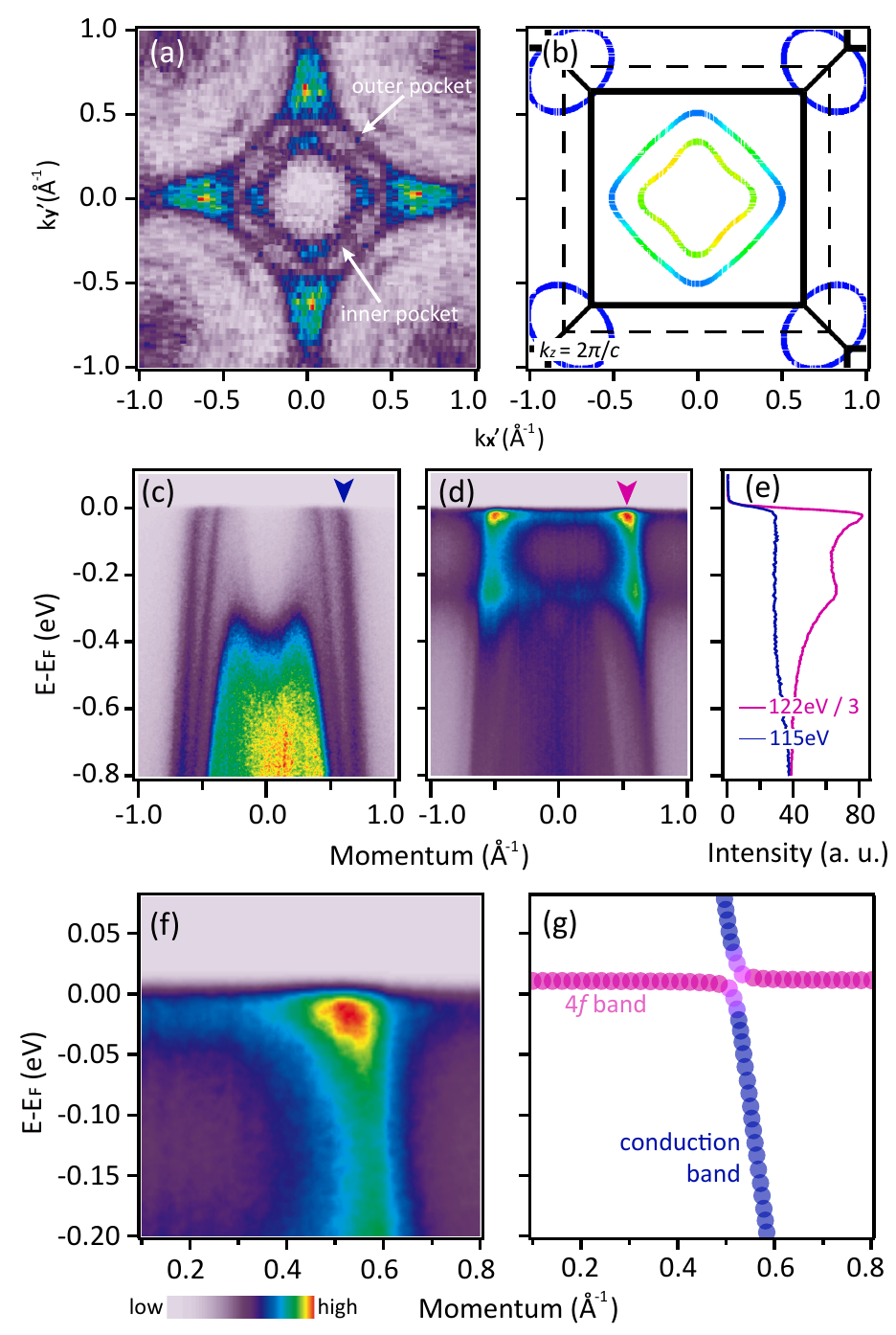}
\caption{(Color Online). 4\textit{f} contribution to the pockets near the bulk \textit{Z} point.  (a) FS near the \textit{Z} point (from the second surface BZ using 150 eV photons). (b) Calculated 2D FS from DFT + \textit{U}, in comparison with (a). The color indicates $v_F$, similar to Fig. 3(b). (c,d,e) Off-resonant (115 eV, c) and on-resonant (122 eV, d) ARPES spectra along $\Bar{\varGamma}$-$\bar{M}$, and the corresponding EDCs at the marked momentum point (e). (f) A zoom-in view of the quasiparticle dispersion near $E_F$. (g) A simulation of the quasiparticle band dispersion using the simplified hybridized band approach within PAM, for comparison with (f).}
\end{figure}

According to RBC and DFT + \textit{U} calculations \cite{ref16,ref18,ref21,ref22,ref24,ref25}, the 4\textit{f} weight mainly resides on the heavy cylindrical band at the \textit{X} point, and the pockets near the \textit{Z} point carry much less 4\textit{f} weight (somtimes ignored). Indeed, the DFT + \textit{U} calculation predicts that the star-shaped pockets near \textit{Z} possess larger band velocity compared to the heavy cylindrical band at \textit{X}, implying smaller 4\textit{f} contribution \cite{ref30}. Fig. 4(a) shows the FS map taken near the \textit{Z} point, which reveals two pockets (inner and outer pockets) centered at $\Bar{\varGamma}$. Here the heavy electron band at $\bar{M}$ appears very broad, due to much weaker intensity compared to other bands. In comparison, the DFT + \textit{U} calculation shows two pockets at $\Bar{\varGamma}$ with the outer pocket exhibiting an intermediate group velocity, implying non-negligible 4\textit{f} weight. To verify the 4\textit{f} contribution, we performed comparison measurements under off-resonant (115 eV) and on-resonant (122 eV) conditions (Fig. 4(c-e)). While the off-resonant spectrum shows sharp non-$4f$ conduction bands, the on-resonant spectrum exhibits clear enhancement of spectral intensity at the Fermi crossings of the conduction bands (marked by an arrow in (d)), signaling clear 4\textit{f} weight. This is best illustrated in the EDCs shown in Fig. 4(e). Note that the resonant enhancement extends to deep energies down to $\sim$-0.4 eV, due to the Kondo process involving the spin-orbit excitation, i.e., the 4$f_{7/2}^1$ peak. A zoom-in view of the quasiparticle dispersion near $E_F$ (Fig. 4(f)) further reveals that the conduction bands exhibit clear bending near $E_F$ and smoothly connect to a weak flat $4f$ band at $\Bar{\varGamma}$, which is characteristic of the \textit{c}-\textit{f} hybridization described by the hybridized band picture within the periodic Anderson model (PAM) \cite{ref33,ref34,ref35}. Within this simplified approach, the non-$4f$ conduction bands (adapted from the localized 4\textit{f} calculation) hybridize with the renormalized 4\textit{f} peaks (the Kondo resonances), yielding a quasiparticle band dispersion similar to a simple two-band mixing. Fig. 4(g) is a simulation based on this model, by setting the hybridization parameter $V$ = 18 meV. The simulation can reasonably explain the experimental data and therefore account for the $4f$ weight observed experimentally. We mention that the quasiparticle dispersion in Fig. 4(f) cannot be well explained by the DFT + \textit{U} calculation, since the conduction band distortions from the 4\textit{f} states are overestimated in DFT + \textit{U}, as discussed earlier. The characteristic band bending as a result of \textit{c}-\textit{f} hybridization, together with the resonance contrast shown in Fig. 4(c-e), provides compelling evidence that the bands near the \textit{Z} point contain appreciable 4\textit{f} weight, which most likely is relevant for the superconductivity in CeCu$_2$Si$_2$.

The above comparison between experiment and calculation shows that the observed FS pockets carry very different 4\textit{f} weight, due to band-dependent \textit{c}-\textit{f} hybridization. The very heavy electron pocket at $\bar{M}$ shows predominant 4\textit{f} character (Fig. 2(a)), which seems to be consistent with RBC (and DFT + \textit{U}, albeit very different effective mass). On the other hand, the hole bands at the \textit{Z} point carry less (albeit still appreciable) 4\textit{f} weight with moderate effective mass due to \textit{c}-\textit{f} hybridization (Fig. 4(f,g)), which is much better explained by the hybridized band picture within PAM. Note that the enclosed area of the outer hole pocket near \textit{Z} is $\sim$0.29 {\AA}$^{-2}$ from Fig. 4(a), close to the values obtained from QO, $\sim$0.28 {\AA}$^{-2}$ in \cite{HuntJPCM1990} and $\sim$0.31 {\AA}$^{-2}$ \cite{TayamaPRB2003}. The observed \textit{c}-\textit{f} hybridization also naturally explains its moderately enhanced effective mass in QO measurements. Although the anisotropic 4\textit{f} distribution in momentum space agrees qualitatively with RBC and DFT + \textit{U} calculations, an accurate and self-consistent description of the overall 4\textit{f} quasiparticles from first-principle calculations is still lacking. While the DFT + DMFT calculation could be better suited to handle the strong local correlation from Ce 4\textit{f} electrons, the existing results do not seem to yield better agreement with experiment (DFT + DMFT calculations appear to show a more isotropic 4\textit{f} weight distribution and incorrect CEF ground state) \cite{ref27,LuoSCPMA2020,ref37}. Our results therefore call for more future studies to understand the fine 4\textit{f} quasiparticles near $E_F$. Note that the Cu atoms in CeCu$_2$Si$_2$ have a partially filled 3\textit{d} shell, and orbital analysis suggests that Cu 3\textit{d} orbitals also contribute to the FS (see Fig. 2(e) and \cite{ref30}). An especially strong Cu $3d$ - Ce $4f$ hybridization had also been conjectured from the extreme pair-breaking capability of tiny substitutions of nonmagnetic (Rh, Pd) as well as magnetic (Mn) impurities on the Cu sites \cite{SpilleHPA}. We mention that the experimental FS of CeCu$_2$Si$_2$ is different from other Ce-based 122 HF compounds, such as CeRu$_2$Ge$_2$ \cite{ref38}, CeRu$_2$Si$_2$ \cite{ref39} and CeRu$_{2}$(Si$_{0.82}$Ge$_{0.18}$)$_{2}$ \cite{ref40}.

Our results provide the electronic basis to understand the unconventional superconductivity in CeCu$_2$Si$_2$, which involves both a sign change of the order parameter and the fully opened gap(s) at very low temperature \cite{ref10,ref16,ref17,ref18}. These superconducting properties share interesting similarity with some Fe-based superconductors, e.g., alkaline Fe-selenides \cite{Park2011PRL,Mou2011PRL,Wang2011EPL,Xu2012PRB,Wang2012EPL}, although the electronic structure of CeCu$_2$Si$_2$ is apparently more complex, consisting of a very heavy quasi-2D band and 3D conduction bands with moderately enhanced mass. In the proposed $d+d$ model \cite{Nica2021npjQM}, the pairing matrix contains a dominating $d_{x^2-y^2}$ intra-band and an additional $d_{xy}$ inter-band component. The former intra-band term relies on pairing within the heavy electron band near \textit{X}, which is observed experimentally in this paper, and it involves singlet pairing across a nesting wave vector spanning the opposite sides of this heavy quasi-2D electron band. The latter could arise from pairing between the heavy electron band near \textit{X} and the moderately heavy hole bands near \textit{Z}, as observed in our experiment. This $d_{xy}$ term had also been conjectured from isothermal in-plane magneto-resistance results \cite{ref15} and is considered to be of fundamental importance for the fully opened gap, see \cite{ref23,Nica2021npjQM}. Therefore, our experimental results provide support for the $d+d$ pairing scenario. Some theoretical calculations suggested that the pairing symmetry after considering the inter-band quantum critical scattering or multipole fluctuation could depend critically on the FS topology \cite{ref21,ref22}. Nevertheless, the alternative pairing symmetries ($s_\pm$ or $s_{++}$) cannot be easily reconciled with the INS results \cite{ref10,ref23}.

In conclusion, we successfully measured the quasiparticle dispersion and FS of the prototypical HF superconductor CeCu$_2$Si$_2$. Our experiments directly reveal the predicted heavy electron band at $E_F$ near the \textit{X} point, and a few other FS bands with appreciable $4f$ contributions. The spectroscopic insight marks an important step towards resolving the HF nature and the enigmatic superconducting state in CeCu$_2$Si$_2$. Future ARPES measurements with higher resolution at sub Kelvin temperatures are needed to resolve the fine structures of the HF bands, as well as the superconducting energy gaps. The method used here could be applicable to other materials that are conventionally difficult for ARPES measurements.

This work is supported by National Key R$\&$D Program of the MOST of China (Grant No. 2017YFA0303100, 2016YFA0300203), National Science Foundation of China (No. 11674280), and the Key R$\&$D Program of Zhejiang Province, China (2021C01002). Part of this research used Beamline 03U of the Shanghai Synchrotron Radiation Facility, which is supported by ME2 project under Contract No. 11227902 from National Natural Science Foundation of China. Y. L. acknowledges useful discussions with Prof. Y.-F. Yang. We would like to thank Dr. Hanna Fedderwitz for help in the synchrotron ARPES measurements.

\end{document}